\documentclass{article}
\pdfoutput=1

\usepackage{etex}
\reserveinserts{12}% 
\usepackage{cmap}
\usepackage[square,numbers]{natbib}
\usepackage[utf8]{inputenc}
\usepackage[T1]{fontenc}
\usepackage[english]{babel}
\usepackage[babel]{csquotes}
\usepackage{microtype,booktabs}
\usepackage{graphicx}
\usepackage{xspace}
\usepackage{multirow}
\usepackage{lmodern}
\usepackage{todonotes}
\usepackage{amsmath,amssymb,amsfonts,amsthm,shuffle}
\usepackage{multirow}
\usepackage{tikz}\usetikzlibrary{arrows,calc}

\usepackage[a4paper,vmargin={3.5cm,3.5cm},hmargin={2.5cm,2.5cm}]{geometry}

\usepackage{enumerate}

\usepackage[breaklinks=true]{hyperref}

\newcommand{\Prob}{\mathbb{P}}
\newcommand{\prob}[1]{\mathbb{P}\left(#1\right)}

\newcommand{\indic}[1]{\mathrm{\textbf{1}}_{#1}}

\newcommand{\Esp}[1]{\mathbb{E}\left[#1\right]}

\newcommand{\setC}{\mathbb{C}}

\newcommand{\setZ}{\mathbb{Z}}

\newcommand{\setR}{\mathbb{R}}

\newcommand{\ca}[1]{\mathcal{#1}}

\newcommand{\ha}[1]{\widehat{#1}}

\newtheorem{definition}{Definition}[section]

\newtheorem{proposition}{Proposition}[section]

\newtheorem{conjecture}{Conjecture}

\usepackage{authblk}

\title{Random knots in 3-dimensional 3-colour percolation: numerical results and conjectures}
\author[1]{Marthe \textsc{de Crouy-Chanel}}
\author[1]{Damien  \textsc{Simon}}
\affil[1]{Sorbonne Université, Laboratoire de probabilités, statistique et modélisation, UMR 8001, 4 place Jussieu, 75005 Paris, France.}

\date{}

\begin{document}
\maketitle
\begin{abstract}
Three-dimensional three-colour percolation on a lattice made of tetrahedra is a direct generalization of two-dimensional two-colour percolation on the triangular lattice. The interfaces between one-colour clusters are made of bicolour surfaces and tricolour non-intersecting and non-self-intersecting curves. Because of the three-dimensional space, these curves describe knots and links. The present paper presents a construction of such random knots using particular boundary conditions and a numerical study of some invariants of the knots. The results are sources of precise conjectures about the limit law of the Alexander polynomial of the random knots.
\end{abstract}

\tableofcontents

\section{Introduction}
\subsection{Context}

There are only a few constructions of random knots in the mathematical literature. The review \cite{modelsofRK} presents various models of random knots with a list of results about them. In general, the construction of random knots either use directly the combinatorial description of knots ---~i.e. a random way of connecting crossings and the random binary choice of the type of crossings or, for another type, the example of \cite{RKinvariants}~---, or the construction of a random non-self-intersecting curve in three-dimension (see \cite{RKpolygon} for example). Our construction is of the second kind.

The relation between knot theory theory and two-dimensional statistical mechanics is quite old now and has helped the construction of various knot invariants, as reviewed in \cite{knotstatmech}. 

The major contribution of the present paper is that the construction of our models of knots is the natural three-dimensional generalization of classical critical two-dimensional percolation model of statistical mechanics, for which many results can be obtained through exact calculations and conformal invariance. The change of dimension from two to three suppresses the conformal invariance but adds very interesting new topological objects, such as knotting. Our study is fully numerical but our results pushed us to formulate precise conjectures. 

The first section presents the model, the analogies and differences with the two-dimensional models and the questions we address. The second section presents the numerical simulations and our results and conjectures.

\paragraph{Acknowledgements.} D.~S. is partially funded by the Grant ANR-14CE25-0014 (ANR GRAAL). We thank Adam Nahum for comments on a first version of the present paper and a nice introduction to the existing results in the physics literature on similar models. We also thank the anonymous referees for additional references and very interesting suggestions.

\subsection{Mathematical definitions}

\paragraph{Geometry.}
For any $N\geq 1$, we call discrete cube of size $N$ the set $C_N=\{0,1,\ldots,N\}^3$ and real cube of size $N$ the set $\ca{C}_N=[0,N]^3$. The real cube $\ca{C}_N$ contains $N^3$ little cubes $c(k_1,k_2,k_3)=[k_1,k_1+1]\times [k_2,k_2+1]\times [k_3,k_3+1]$ where $0\leq k_1,k_2,k_3<N$ are integers.

\begin{figure}
\begin{center}
\begin{tikzpicture}
\tikzstyle{scopeangle}=[x={(1cm,0)}, y={(0.6cm,0.3cm)}, z={(0,1cm)}]
\begin{scope}[scopeangle,ultra thick]
\coordinate (a000) at (0,0,0); 
\coordinate (a100) at (1,0,0);
\coordinate (a010) at (0,1,0);
\coordinate (a001) at (0,0,1);
\coordinate (a110) at (1,1,0);
\coordinate (a101) at (1,0,1);
\coordinate (a011) at (0,1,1);
\coordinate (a111) at (1,1,1);
\draw (a000) -- (a100) -- (a110) ;
\draw[dashed] (a110) -- (a010) -- (a000);
\draw (a001) -- (a101) -- (a111);
\draw (a111) -- (a011) -- (a001);
\draw (a000) -- (a001);
\draw (a100) -- (a101);
\draw[dashed] (a010) -- (a011);
\draw (a110) -- (a111);
\draw[->,thin] (a000) -- (+2,0,0) node [above]{$x$};
\draw[->,thin] (a000) -- (+0,2,0) node [right] {$y$};
\draw[->,thin] (a000) -- (+0,0,2) node [right] {$z$};
\end{scope}

//(1,0,0) (1,1,0)
\begin{scope}[ultra thick, xshift=3cm, yshift=-1cm,scopeangle]
\coordinate (a000) at (0,0,0);
\coordinate (a100) at (1,0,0);
\coordinate (a010) at (0,1,0);
\coordinate (a001) at (0,0,1);
\coordinate (a110) at (1,1,0);
\coordinate (a101) at (1,0,1);
\coordinate (a011) at (0,1,1);
\coordinate (a111) at (1,1,1);
\draw[dotted,thin] (a000) -- (a100) -- (a110) ;
\draw[dotted,thin] (a110) -- (a010) -- (a000);
\draw[dotted,thin] (a001) -- (a101) -- (a111) -- (a011) -- (a001);
\draw[dotted,thin] (a000) -- (a001);
\draw[dotted,thin] (a100) -- (a101);
\draw[dotted,thin] (a010) -- (a011);
\draw[dotted,thin] (a110) -- (a111);

\draw (a000) -- (a100) -- (a110) -- (a111) -- (a000);
\draw[dashed] (a000) -- (a110);
\draw (a100) -- (a111);
\end{scope}

//(1,0,0) (1,0,1)
\begin{scope}[ultra thick, xshift=3cm, yshift=1cm,scopeangle]
\coordinate (a000) at (0,0,0);
\coordinate (a100) at (1,0,0);
\coordinate (a010) at (0,1,0);
\coordinate (a001) at (0,0,1);
\coordinate (a110) at (1,1,0);
\coordinate (a101) at (1,0,1);
\coordinate (a011) at (0,1,1);
\coordinate (a111) at (1,1,1);
\draw[dotted,thin] (a000) -- (a100) -- (a110) ;
\draw[dotted,thin] (a110) -- (a010) -- (a000);
\draw[dotted,thin] (a001) -- (a101) -- (a111) -- (a011) -- (a001);
\draw[dotted,thin] (a000) -- (a001);
\draw[dotted,thin] (a100) -- (a101);
\draw[dotted,thin] (a010) -- (a011);
\draw[dotted,thin] (a110) -- (a111);

\draw (a000) -- (a100) -- (a101) -- (a111);
\draw[dashed] (a000) -- (a111);
\draw (a000) -- (a101);
\draw (a100) -- (a111);
\end{scope}

//(0,1,0) (1,1,0)
\begin{scope}[ultra thick, xshift=5cm, yshift=-1cm,scopeangle]
\coordinate (a000) at (0,0,0);
\coordinate (a100) at (1,0,0);
\coordinate (a010) at (0,1,0);
\coordinate (a001) at (0,0,1);
\coordinate (a110) at (1,1,0);
\coordinate (a101) at (1,0,1);
\coordinate (a011) at (0,1,1);
\coordinate (a111) at (1,1,1);
\draw[dotted,thin] (a000) -- (a100) -- (a110) ;
\draw[dotted,thin] (a110) -- (a010) -- (a000);
\draw[dotted,thin] (a001) -- (a101) -- (a111) -- (a011) -- (a001);
\draw[dotted,thin] (a000) -- (a001);
\draw[dotted,thin] (a100) -- (a101);
\draw[dotted,thin] (a010) -- (a011);
\draw[dotted,thin] (a110) -- (a111);

\draw[dashed] (a000) -- (a010) -- (a110);
\draw (a110)-- (a111)-- (a000);
\draw (a000) -- (a110);
\draw[dashed] (a010) -- (a111);
\end{scope}

//(0,1,0) (0,1,1)
\begin{scope}[ultra thick, xshift=5cm, yshift=1cm,scopeangle]
\coordinate (a000) at (0,0,0);
\coordinate (a100) at (1,0,0);
\coordinate (a010) at (0,1,0);
\coordinate (a001) at (0,0,1);
\coordinate (a110) at (1,1,0);
\coordinate (a101) at (1,0,1);
\coordinate (a011) at (0,1,1);
\coordinate (a111) at (1,1,1);
\draw[dotted,thin] (a000) -- (a100) -- (a110) ;
\draw[dotted,thin] (a110) -- (a010) -- (a000);
\draw[dotted,thin] (a001) -- (a101) -- (a111) -- (a011) -- (a001);
\draw[dotted,thin] (a000) -- (a001);
\draw[dotted,thin] (a100) -- (a101);
\draw[dotted,thin] (a010) -- (a011);
\draw[dotted,thin] (a110) -- (a111);

\draw (a000) -- (a010);
\draw[dashed] (a010) -- (a011);
\draw (a011) -- (a111) -- (a000);
\draw (a000) -- (a011);
\draw (a010) -- (a111);
\end{scope}

//(0,0,1) (0,1,1)
\begin{scope}[ultra thick, xshift=7cm, yshift=-1cm,scopeangle]
\coordinate (a000) at (0,0,0);
\coordinate (a100) at (1,0,0);
\coordinate (a010) at (0,1,0);
\coordinate (a001) at (0,0,1);
\coordinate (a110) at (1,1,0);
\coordinate (a101) at (1,0,1);
\coordinate (a011) at (0,1,1);
\coordinate (a111) at (1,1,1);
\draw[dotted,thin] (a000) -- (a100) -- (a110) ;
\draw[dotted,thin] (a110) -- (a010) -- (a000);
\draw[dotted,thin] (a001) -- (a101) -- (a111) -- (a011) -- (a001);
\draw[dotted,thin] (a000) -- (a001);
\draw[dotted,thin] (a100) -- (a101);
\draw[dotted,thin] (a010) -- (a011);
\draw[dotted,thin] (a110) -- (a111);

\draw (a000) -- (a001) -- (a011) -- (a111) -- (a000);
\draw[dashed] (a000) -- (a011);
\draw (a001) -- (a111);
\end{scope}

//(0,0,1) (1,0,1)
\begin{scope}[ultra thick, xshift=7cm, yshift=1cm,scopeangle]
\coordinate (a000) at (0,0,0);
\coordinate (a100) at (1,0,0);
\coordinate (a010) at (0,1,0);
\coordinate (a001) at (0,0,1);
\coordinate (a110) at (1,1,0);
\coordinate (a101) at (1,0,1);
\coordinate (a011) at (0,1,1);
\coordinate (a111) at (1,1,1);
\draw[dotted,thin] (a000) -- (a100) -- (a110) ;
\draw[dotted,thin] (a110) -- (a010) -- (a000);
\draw[dotted,thin] (a001) -- (a101) -- (a111) -- (a011) -- (a001);
\draw[dotted,thin] (a000) -- (a001);
\draw[dotted,thin] (a100) -- (a101);
\draw[dotted,thin] (a010) -- (a011);
\draw[dotted,thin] (a110) -- (a111);

\draw (a000) -- (a001) -- (a101) -- (a111);
\draw (a111) -- (a000);
\draw (a000) -- (a101);
\draw (a001) -- (a111);
\end{scope}
\end{tikzpicture}
\end{center}
\caption{\label{fig:tetrahedra}Canonical splitting of a unit cube into six tetrahedra.}
\end{figure}
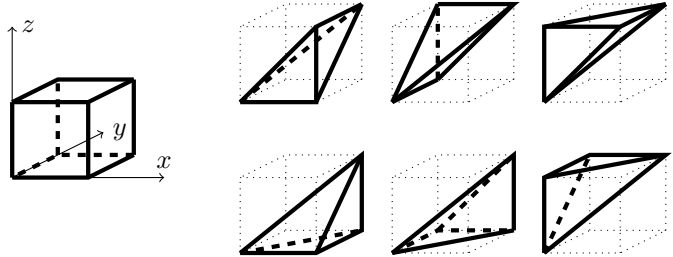

Each cube $c(k_1,k_2,k_3)$ can be divided into six tetrahedra $t_{\rho}(k_1,k_2,k_3)$ where $\rho$ is a permutation of $\{1,2,3\}$ such that \begin{equation}\label{eq:deftetrahedron}
t_{\rho}(k_1,k_2,k_3)=\{ (k_1+x_1,k_2+x_2,k_3+x_3); \quad  (x_1,x_2,x_3)\in[0,1]^3 \quad 0\leq x_{\rho(1)} \leq x_{\rho(2)}\leq x_{\rho(3)} \leq 1 \}
\end{equation}
as represented in figure~\ref{fig:tetrahedra}.
The faces (resp. edges, vertices) of the tetrahedron are the subsets of $t_{\rho}(k_1,k_2,k_3)$ such that one (resp. two, three) out of the four inequalities in \eqref{eq:deftetrahedron} is (resp. are, are) replaced by an equality. 

%The set of the tetrahedra (resp. faces, resp. edges) of a real cube $\ca{C}_N$ is written $T_N$ (resp. $F_N$, resp. $E_N$).

\paragraph{Percolation.}

\begin{definition}[3-colour percolation with free boundary condition]
For any $N\geq 1$, a 3-colour percolation model with free boundary condition on the cube $C_N$ is a family of random variables $(U_x)_{x\in C_N}$ with values in $\{0,1,2\}$ such that:
\begin{enumerate}[(i)]
\item the $(U_x)$ are independent
\item for every $x\in C_N$, the law of $U_x$ is uniform on $\{0,1,2\}$.
\end{enumerate}
\end{definition}

\begin{definition}[3-colour percolation with Dobrushin boundary condition]\label{def:dobrushinperco}
For any $N\geq 1$, a 3-colour percolation model with Dobrushin boundary condition on the cube $C_N$ is a family of random variables $(U_x)_{x\in C_N}$ with values in $\{0,1,2\}$ such that:
\begin{enumerate}[(i)]
\item the $(U_x)_{x\in C_N}$ are independent
\item for every $x\in \{1,\ldots,N-1\}^3$, the law of $U_x$ is uniform on $\{0,1,2\}$.
\item for $x\in \{(0,0,0),(0,N,0),(0,0,N),(N,N,0),(0,N,N),(N,N,N)\} \cup F^{(1)}_0 \cup F^{(2)}_N \cup E^{(2,3)}_{N,0} \cup E^{(2,3)}_{N,N} \cup E^{(1,3)}_{0,N} \cup E^{(1,2)}_{0,0} \cup E^{(1,2)}_{0,N} \cup E^{(1,2)}_{N,N} $, $U_x=0$ a.s.,
\item for $x\in \{(N,0,N)\}\cup F^{(2)}_0 \cup F^{(3)}_N \cup E^{(2,3)}_{0,N} \cup E^{(1,3)}_{N,N} $, $U_x=1$ a.s.,
\item for $x\in \{(N,0,0)\}\cup F^{(1)}_N \cup F^{(3)}_0 \cup E^{(2,3)}_{0,0} \cup E^{(1,3)}_{0,0} \cup E^{(1,3)}_{N,0} \cup E^{(1,2)}_{N,0}$, $U_x=2$ a.s.
\end{enumerate}
where $F^{(i)}_k$ is the strict face of the cube which the $i$-th coordinate is equal to $k$ and $E^{(i,j)}_{k,l}$ is the strict edge of the cube for which the $i$-th coordinate is equal to $k$ and the $j$-coordinate is equal to $l$.
\end{definition}

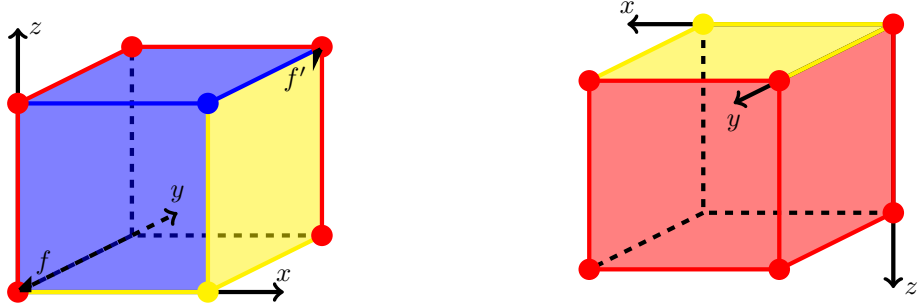
\begin{figure}
\begin{center}
\begin{tikzpicture}
\tikzstyle{scopeangle}=[x={(1cm,0)}, y={(0.6cm,0.3cm)}, z={(0,1cm)}]
\tikzstyle{color0}=[red]
\tikzstyle{color1}=[blue]
\tikzstyle{color2}=[yellow]
\begin{scope}[scopeangle,ultra thick,scale=2.5]
\coordinate (a000) at (0,0,0); 
\coordinate (a100) at (1,0,0);
\coordinate (a010) at (0,1,0);
\coordinate (a001) at (0,0,1);
\coordinate (a110) at (1,1,0);
\coordinate (a101) at (1,0,1);
\coordinate (a011) at (0,1,1);
\coordinate (a111) at (1,1,1);
% DEVANT:
\draw[dashed] (a110) -- (a010); % E^23_N0
\draw[dashed] (a010) -- (a000); % E^13_00
\draw[dashed] (a010) -- (a011); % E^12_0N
\fill[color1,opacity=0.5] (a000) -- (a100) -- (a101) -- (a001) --cycle;
\fill[color1,opacity=0.5] (a001) -- (a011) -- (a111) -- (a101) --cycle;
\fill[color2,opacity=0.5] (a100) -- (a110) -- (a111) -- (a101) --cycle;
% DERRIERE:
%\fill[color0,opacity=0.5] (a000) -- (a010) -- (a011) -- (a001) --cycle;
%\fill[color0,opacity=0.5] (a010) -- (a011) -- (a111) -- (a110) --cycle;
%\fill[color2,opacity=0.5] (a000) -- (a010) -- (a110) -- (a100) --cycle;
\draw[->] (a000) -- (+1.4,0,0) node [above]{$x$};
\draw[->,dashed] (a000) -- (+0,1.4,0) node [above] {$y$};
\draw[->] (a000) -- (+0,0,1.4) node [right] {$z$};
\draw[color2,ultra thick] (a000) -- (a100) -- (a110) ; % E^23_00 et E^13_N0
\draw[color1,ultra thick] (a001) -- (a101); % E^23_0N
\draw[color1,ultra thick] (a101) -- (a111); %E^13_NN
\draw[color0,ultra thick] (a111) -- (a011); %E^23_NN
\draw[color0,ultra thick] (a011) -- (a001); %E^13_0N
\draw[color0,ultra thick] (a000) -- (a001); %E^12_00
\draw[color2,ultra thick] (a100) -- (a101); %E^12_N0
\draw[color0,ultra thick] (a110) -- (a111); %E^12_NN
\node at (a000) [fill,color0,circle,inner sep=0.1cm] {};
\node at (a100) [fill,color2,circle,inner sep=0.1cm] {};
%\node at (a010) [fill,black,circle,inner sep=0.1cm] {};
\node at (a001) [fill,color0,circle,inner sep=0.1cm] {};
\node at (a110) [fill,color0,circle,inner sep=0.1cm] {};
\node at (a101) [fill,color1,circle,inner sep=0.1cm] {};
\node at (a011) [fill,color0,circle,inner sep=0.1cm] {};
\node at (a111) [fill,color0,circle,inner sep=0.1cm] {};

\filldraw[black] (0.01,0.0,0.01) -- (0.06,0.0,0.01) -- (0.06,0.0,0.06) -- (0.01,0.0,0.01); 
\node at (0.035,0,0.035) [above right] {$f$};
\filldraw[black] (1.,0.99,0.99) -- (1.,0.94,0.99) -- (1,0.94,0.94) -- (1,0.99,0.99); 
\node at (1.,0.965,0.965) [below left] {$f'$};
\end{scope}
\end{tikzpicture}
\hspace{3cm}
\begin{tikzpicture}
\tikzstyle{scopeangle}=[x={(-1cm,0)}, y={(-0.6cm,-0.3cm)}, z={(0,-1cm)}]
\tikzstyle{color0}=[red]
\tikzstyle{color1}=[blue]
\tikzstyle{color2}=[yellow]
\begin{scope}[scopeangle,ultra thick,scale=2.5]
\coordinate (a000) at (0,0,0); 
\coordinate (a100) at (1,0,0);
\coordinate (a010) at (0,1,0);
\coordinate (a001) at (0,0,1);
\coordinate (a110) at (1,1,0);
\coordinate (a101) at (1,0,1);
\coordinate (a011) at (0,1,1);
\coordinate (a111) at (1,1,1);
% DEVANT:
%\fill[color1,opacity=0.5] (a000) -- (a100) -- (a101) -- (a001) --cycle;
%\fill[color1,opacity=0.5] (a001) -- (a011) -- (a111) -- (a101) --cycle;
%\fill[color2,opacity=0.5] (a100) -- (a110) -- (a111) -- (a101) --cycle;
% DERRIERE:
\fill[color0,opacity=0.5] (a000) -- (a010) -- (a011) -- (a001) --cycle;
\fill[color0,opacity=0.5] (a010) -- (a011) -- (a111) -- (a110) --cycle;
\fill[color2,opacity=0.5] (a000) -- (a010) -- (a110) -- (a100) --cycle;
\draw[->] (a000) -- (+1.4,0,0) node [above]{$x$};
\draw[->] (a000) -- (+0,1.4,0) node [below] {$y$};
\draw[->] (a000) -- (+0,0,1.4) node [right] {$z$};
\draw[color2,ultra thick] (a000) -- (a100) -- (a110) ; % E^23_00 et E^13_N0
\draw[dashed] (a001) -- (a101); % E^23_0N
\draw[dashed] (a101) -- (a111); %E^13_NN
\draw[color0,ultra thick] (a111) -- (a011); %E^23_NN
\draw[color0,ultra thick] (a011) -- (a001); %E^13_0N
\draw[color0,ultra thick] (a000) -- (a001); %E^12_00
\draw[dashed] (a100) -- (a101); %E^12_N0
\draw[color0,ultra thick] (a110) -- (a111); %E^12_NN
\draw[color0,ultra thick] (a110) -- (a010); % E^23_N0
\draw[color2,ultra thick] (a010) -- (a000); % E^13_00
\draw[color0,ultra thick] (a010) -- (a011); % E^12_0N

\node at (a000) [fill,color0,circle,inner sep=0.1cm] {};
\node at (a100) [fill,color2,circle,inner sep=0.1cm] {};
\node at (a010) [fill,color0,circle,inner sep=0.1cm] {};
\node at (a001) [fill,color0,circle,inner sep=0.1cm] {};
\node at (a110) [fill,color0,circle,inner sep=0.1cm] {};
%\node at (a101) [fill,black,circle,inner sep=0.1cm] {};
\node at (a011) [fill,color0,circle,inner sep=0.1cm] {};
\node at (a111) [fill,color0,circle,inner sep=0.1cm] {};
\end{scope}
\end{tikzpicture}
\end{center}
\caption{\label{fig:dobrushin} Dobrushin boundary conditions from definition~\ref{def:dobrushinperco}: two views of the same cube in order to see the colours of all the faces, edges and vertices. Red corresponds to $0$, blue to $1$ and yellow to $2$.}
\end{figure}

The boundary conditions are represented on figure~\ref{fig:dobrushin}. This is the generalization with three dimensions and three colours of the classical critical percolation with two dimensions and two colours on the triangular lattice. This observation explains many features of our construction presented in the next section.

One checks in the previous definition that the only two tricolour faces on the boundary of the cube are\begin{align}\label{eq:tricolorfacesboundary}
f&=\{(0,0,0),(1,0,0),(1,0,1)\} 
\\ 
f'&=\{(N,N,N), (N,N-1,N), (N,N-1,N-1).
\end{align}

\subsection{Geometrical structure of the model}

\paragraph{Two dimensions and two colours.}
As explained in the review \cite{smirnovicm} for two-dimensional two-colour percolation, the triangular lattice plays a special role since it allows for a well-defined notion  of boundary of monochrome clusters. If a triangular face contains both colours on its vertices, one colour is present twice and the other is present once: the interface between the two colours is defined as the segment that joins the middles of the two bicolour edges. 

Interfaces between two colours are thus non-intersecting and non-self-intersecting loops inside the domain, if the boundary is monochrome. On the other hand, if the boundary of the domain is made of one loop obtained as the concatenation of two monochrome curve -- this situation is known as a Dobrushin boundary condition --, the interfaces are internal loops together with an additional non-self-intersecting curve that connects the two edges of the boundary that corresponds to a change of colour.

\paragraph{Three dimensions and three colours.}

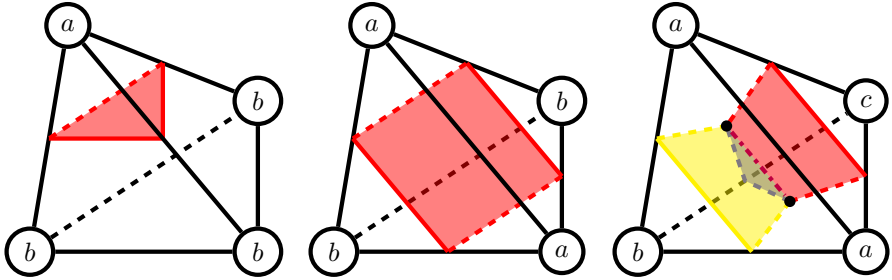
\begin{figure}
\begin{center}
\begin{tikzpicture}
% 3 fois a, 1 fois b
\begin{scope}[scale=0.5,ultra thick]
\node (A1) at (-1,0) [circle,draw] {$b$};%bas gauche
\node (A2) at (0,6) [circle,draw] {$a$};%haut gauche
\node (A3) at (5,0) [circle,draw] {$b$};%bas droit
\node (A4) at (5,4) [circle,draw] {$b$};% haut droit
\draw (A1) -- (A2);
\draw (A1) -- (A3);
\draw[dashed] (A1) -- (A4);
\draw (A2) -- (A4);
\draw (A3) -- (A4);
\coordinate (m12) at ($(A1)!0.5!(A2)$) ;
\coordinate (m13) at ($(A1)!0.5!(A3)$) ;
\coordinate (m14) at ($(A1)!0.5!(A4)$) ;
\coordinate (m23) at ($(A2)!0.5!(A3)$);
\coordinate (m24) at ($(A2)!0.5!(A4)$);
\coordinate (m34) at ($(A3)!0.5!(A4)$);
%\coordinate (f124) at ($(m12)!0.3333!(A4)$);
%\coordinate (f134) at ($(m13)!0.3333!(A4)$);
\fill[color=red,opacity=0.5] (m12) -- (m23) -- (m24) -- cycle; 
\draw[color=red,ultra thick] (m12) -- (m23) -- (m24) ; 
\draw[color=red,ultra thick, dashed] (m24) -- (m12) ; 
\draw (A2) -- (A3);
\end{scope}
% 2 fois a, 2 fois b
\begin{scope}[scale=0.5,xshift=8.cm,ultra thick]
\node (A1) at (-1,0) [circle,draw] {$b$};%bas gauche
\node (A2) at (0,6) [circle,draw] {$a$};%haut gauche
\node (A3) at (5,0) [circle,draw] {$a$};%bas droit
\node (A4) at (5,4) [circle,draw] {$b$};% haut droit
\draw (A1) -- (A2);
\draw (A1) -- (A3);
\draw[dashed] (A1) -- (A4);

\draw (A2) -- (A4);
\draw (A3) -- (A4);
\coordinate (m12) at ($(A1)!0.5!(A2)$) ;
\coordinate (m13) at ($(A1)!0.5!(A3)$) ;
\coordinate (m14) at ($(A1)!0.5!(A4)$) ;
\coordinate (m23) at ($(A2)!0.5!(A3)$);
\coordinate (m24) at ($(A2)!0.5!(A4)$);
\coordinate (m34) at ($(A3)!0.5!(A4)$);
%\coordinate (f124) at ($(m12)!0.3333!(A4)$);
%\coordinate (f134) at ($(m13)!0.3333!(A4)$);
\fill[color=red,opacity=0.5] (m12) -- (m24) -- (m34) -- (m13) -- cycle; 
\draw[color=red,ultra thick, dashed] (m12) -- (m24);
\draw[color=red,ultra thick] (m24) -- (m34);
\draw[color=red,ultra thick, dashed] (m34) -- (m13);
\draw[color=red,ultra thick] (m13) -- (m12) ;
\draw (A2) -- (A3);
\end{scope}
% 2 fois a, 1 fois b, 1 fois c
\begin{scope}[scale=0.5,xshift=16.cm,yshift=0.cm, ultra thick]
\node (A1) at (-1,0) [circle,draw] {$b$};%bas gauche
\node (A2) at (0,6) [circle,draw] {$a$};%haut gauche
\node (A3) at (5,0) [circle,draw] {$a$};%bas droit
\node (A4) at (5,4) [circle,draw] {$c$};% haut droit
\draw (A1) -- (A2);
\draw (A1) -- (A3);
\draw[dashed] (A1) -- (A4);

\draw (A2) -- (A4);
\draw (A3) -- (A4);
\coordinate (m12) at ($(A1)!0.5!(A2)$) ;
\coordinate (m13) at ($(A1)!0.5!(A3)$) ;
\coordinate (m14) at ($(A1)!0.47!(A4)$) ;
\coordinate (m23) at ($(A2)!0.5!(A3)$);
\coordinate (m24) at ($(A2)!0.5!(A4)$);
\coordinate (m34) at ($(A3)!0.5!(A4)$);
\coordinate (f124) at ($(m12)!0.3333!(A4)$) ;
\coordinate (f134) at ($(m13)!0.3333!(A4)$) ;
\fill[color=blue,opacity=0.5] (f124)--(f134) -- (m14) --  (f124); 
\draw[color=blue,ultra thick, dashed]  (f134) -- (m14) --  (f124); 
\fill[color=red,opacity=0.5] (f124) -- (f134) -- (m34) -- (m24) -- (f124); 
\draw[color=red,ultra thick, dashed] (m24) -- (f124) ;
\draw[color=red,ultra thick, dashed] (f134) -- (m34);
\draw[color=red,ultra thick] (m34)-- (m24); 
\fill[color=yellow,opacity=0.5] (f124) -- (f134) -- (m13) -- (m12) -- (f124); 
\draw[color=yellow,ultra thick, dashed]  (f134) -- (m13);
\draw[color=yellow,ultra thick, dashed] (m12) -- (f124); 
\draw[color=yellow,ultra thick] (m12) -- (m13); 
\node (fn124) at ($(m12)!0.3333!(A4)$) [circle,fill,inner sep=1.5pt] {};
\node (fn134) at ($(m13)!0.3333!(A4)$) [circle,fill,inner sep=1.5pt] {};
\draw[ultra thick,purple,dashdotted] (fn124) -- (fn134);

\draw (A2) -- (A3);
\end{scope}
\end{tikzpicture}
\end{center}
\caption{\label{fig:clusterboundarygeometry} Possible configurations on a tetrahedron and cluster interfaces. The first two cases correspond to the elements of surfaces (in red/gray) separating two colours $a$ and $b$. The third case corresponds to a 3-colour line (thick purple line), on which three elements of surfaces meet (a triangle between $b$ and $c$ and two quadrilaterals, one between $a$ and $b$ and one between $a$ and $c$).}
\end{figure}

\emph{Mutatis mutandis}, all the previous properties can be generalized easily to the case of three dimensions and three colours. Triangles become tetrahedra and the triangular lattice becomes the three-dimensional lattice described by the tetrahedra of \eqref{eq:deftetrahedron}. The interfaces between colours are again well-defined except that they are now made of 2-colour surfaces and 3-colour lines, as explained below and in figure \ref{fig:clusterboundarygeometry}. There are two types of tetrahedra with two colours:
\begin{itemize}
\item one colour is represented once and the other thrice: the interface is defined as the triangle that joins the middles of the three bicoloured edges.
\item both colours are represented twice: the interface is defined as the quadrilateral that joins the four bicoloured edges.
\end{itemize}
There is only one type of tetrahedra with three colours: one colour, let say $a$, is represented twice and the two others, let say $b$ and $c$, are represented once. There are two faces with three colours, whose centres are $m_1$ and $m_2$; there is one edge with $b$ and $c$, one edge with both $a$, two edges with $a$ and $b$ and two with $a$ and $c$. We now define the interfaces as follows:
\begin{itemize}
\item the interface between $a$ and $b$ (resp. $c$) is the quadrilateral that joins the middles of the edges with $a$ and $b$ (resp. $c$) and the two points $m_1$ and $m_2$;
\item the interface between $b$ and $c$ is the triangle that joins the middle of the edge with $b$ and $c$ and the two points $m_1$ and $m_2$. 
\item the three elements of surface join on the segment that joins $m_1$ and $m_2$, which is an element of a $3$-colour line. 
\end{itemize}

A bicolour surface is thus a surface made of quadrilaterals and triangles, which may have or not a boundary. If it has a boundary, it is made of non-self-intersecting loops that either belong to the boundary of the domain or are (part of) a $3$-colour line. The set of $3$-colour lines forms a set of non-intersecting and non-self-intersecting curves that are either loops or curves with extremities on the boundary of the domain.

\begin{figure}\begin{center}
\input{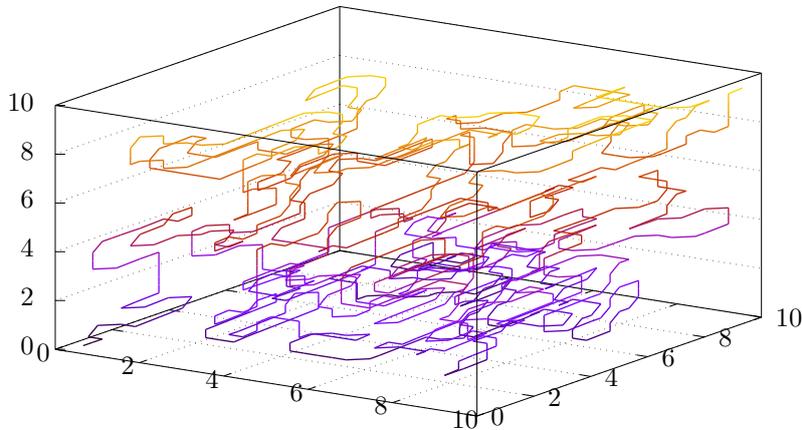}
\end{center}
\caption{\label{fig:knotexample} Example of a random curve $\Gamma$ in a cube of size $N=10$ with Dobrushin boundary conditions. Any  curve $\gamma$ that remains in the boundary of the cube and joins the extremities of $\Gamma$ defines by concatenation with $\Gamma$ the same knot.}
\end{figure}

\begin{proposition}
A 3-colour percolation model with Dobrushin boundary condition on the cube $C_N$ defines a random non-self-intersecting piecewise affine curve $\Gamma:[0,L_N]\to \ca{C}_N$ as the unique $3$-colour line made of $L_N$ segments that joins the centres of the face $f$ and $f'$ defined in eq.~\eqref{eq:tricolorfacesboundary}.
\end{proposition}
The proof is easily deduced from the previous geometric considerations and generalizes the two-dimensional construction with two colours. An example is presented in figure~\ref{fig:knotexample}.

\paragraph{Knots.} Two non-self-intersecting continuous loops $\gamma_0,\gamma_1: [0,1] \to \setR^3$ (with $\gamma_i(0)=\gamma_i(1)$ since these are loops) are equivalent if there exists a continuous function $h:[0,1]\times [0,1]\to\setR^3$ such that $h(t,0)=\gamma_0(t)$, $h(t,1)=\gamma_1(t)$ and, for any $s\in[0,1]$, $t\mapsto h(t,s)$ is injective excepted for $h(0,s)=h(1,s)$. An equivalence class is called a knot.

Knot theory is all about the description of these equivalence classes and deciding whether two non-self-intersecting continuous loops are equivalent. A knot is tame if one of its representing loop is locally flat or $C^\infty$, or piecewise affine. The possibility to choose between these various notions of smoothness is already a difficult result. A knot which is not tame is said to be wild. Most of the time (and in all the constructions discussed below), the constructions made in knot theory are in fact made for tame knots. This is in particular the case for the knots we study in our discrete models (but not necessarily for their scaling limit!). The interested reader may consult \cite{lickorish} for a nice and easy-to-read introduction to knot theory and knot invariants with a good bibliography.

\begin{proposition}
Let $\gamma$ be any piecewise affine curve that remains in the boundary of the cube $\partial \ca{C}_N$, joining the centres of the faces $f$ and $f'$ of eq.~\eqref{eq:tricolorfacesboundary}.

A 3-colour percolation model with Dobrushin boundary condition on the cube $C_N$ defines a random knot $K$ as the equivalence class of the curve $\Gamma$ closed with the curve $\gamma$ that joins its extremities; the construction of $K$ does not depend on the particular choice of $\gamma$.
\end{proposition}

Up to our knowledge, this is the first time that such knot considerations emerge in three-dimensional percolation-type model. The construction based on three colours on a well-chosen lattice has already appeared in \cite{sheffieldyadin} in the mathematical literature and in \cite{bradleyetal} in the physical literature but the authors only considered the fractal and metric properties of the $3$-colour line. 

\subsection{Addressed questions.}

\paragraph{Dimension $2$ versus dimension $3$.}
Critical two-dimensional two-colour percolation on the triangular lattice is known to be described by conformal field theory \cite{smirnovicm}. In particular, the interfaces are known to be described by conformal loop ensembles and the interface particularized by Dobrushin boundary conditions is known to be described by a Schramm-Loewner Evolution curve $SLE(6)$. In our case, no such conformal property is expected to be present; however, fractal behaviours for $L_N$ are still expected as for any critical model.

A family of non-intersecting and non-self-intersecting loops in two dimensions has a simple topology: each loop is equivalent to a circle and divides the plane into two domains, one interior and one exterior. A second loop belongs either to the interior domain of a first loop or to its exterior. In three dimensions, the topology is much more involved: a single loop may form a non-trivial knot and two loops may form a non-trivial link. Up to our knowledge, no study of such knots arising in some simple probabilistic model on $\setZ^3$ has been previously performed and we started simulations without expecting any particular behaviour and having any prediction.

\paragraph{Knot structure and identification of knots.} Knots, defined as above as equivalence classes of three-dimensional non-self-intersecting curves, are difficult to identify since it is hard, from an algorithmic point of view, to detect whether two curves define the same knot. Thus, since our model produces large knots, we have not been able to fully identify these knots. 

However, knots have \emph{arithmetic properties} and can be classified by \emph{invariants}. As described in \cite{lickorish}, there are notions of product of two knots, of prime knots, of decomposition into prime knots. Moreover, one can build knot invariants, as objects computed on curves that are constant on an equivalence class. Relevant invariants for our study are those that take values in rings such that integers or polynomials and are compatible with the arithmetic properties of knots: if  a knot $K$ is the product of $K_1$ and $K_2$, the invariants we use are such that $I(K)=I(K_1)\cdot I(K_2)$. The divisibility properties of invariants are good heuristic hints about the structure of a knot but are not clear identification of the knots.

We focus in the next section on two invariants that are integer-valued and cheap to compute, so that we can accumulate large statistics. These two invariants are the absolute values of the Alexander polynomial of a knot at the points $-1$ and $i$. Computing the Alexander  polynomial $\Delta_K(t)$ of a knot $K$ up to a factor $\pm t^k$ where $k\in\setZ$ requires only the evaluation of the determinant of a sparse matrix. Taking the absolute value and focusing on $-1$ and $i$ makes it unnecessary to identify the factor $\pm t^k$. Moreover the properties of $\Delta_K(t)$ imply that the two values are integers and this allows us to control the numerical inaccuracy in our computations.
These restrictions are not enough to identify and factorize knots but, as we will see below, the results already give nice indications about the structure of the random knots.

\section{Numerical simulations and results}
\subsection{Mathematical description of the simulations}

Our simulations are made of several steps and use various mathematical tricks that we present here. The colours $(U_x)_{x\in C_N}$ inside the cube are generated sequentially from a Mersenne twister random number generator. The $3$-colour curve is then extracted by following the tetrahedra with $3$-colour, entering by the $3$-colour face $f$ and exiting by the other one $f'$. 

The idea is then to project the curve on a plane, collect its self-crossings while keeping the information of which part was above or below before projection and build the combinatorial structure of the knot.

\paragraph{Projection on a plane: a random trick.}
In order to avoid numerical computations on the coordinates of the $3$-colour line, we choose to project the three-dimensional curve on the plane $z=0$. However, using the center of the $3$-coloured faces of tetrahedra to define the $3$-colour line leads to high degeneracy of the projected curve, for which many segments are fully identified. In order to avoid this degeneracy, we define the extremities of the segments of a $3$-colour line as a random positive barycentre of the three vertices of the corresponding $3$-colour face, such that, almost surely, no projections overlap while the knot remains unchanged.

\paragraph{Combinatorial structure inspired by planar maps.}

We store a knot in a computer as two permutations ($\sigma$ and $\alpha$ described below) and a function ($\tau$ described below) on a set of $4n$ elements as described below. This coding is inspired by the coding of combinatorial planar maps, to which we add a height information.
\begin{definition}[encoding of a knot]\label{def:knotencoding}
The encoding of a knot is a finite set $H_{4n}=\{0,1,\ldots,4n-1\}$ with three permutations $\sigma$,$\alpha$,$\phi$ on $H_{4n}$ and a function $\tau: H_{4n} \to \{-1,1\}$ such that:
\begin{enumerate}[(i)]
\item $\phi\circ\alpha\circ \sigma= id$;
\item the group generated by $\sigma$, $\alpha$ and $\phi$ acts transitively on $H_{4n}$;
\item all cycles of $\alpha$ have length $2$;
\item all cycles of $\sigma$ have length $4$;
\item all cycles of $\sigma^2\circ\alpha$ have length $2n$;
\item for all $u\in H_{4n}$, $\tau(\sigma(u))=-\tau(u)$.
\end{enumerate}
\end{definition}
A example on the trefoil knot is represented in figure~\ref{fig:knotencoding}. Each element of $H_{4n}$ corresponds to one of the four directions that go out of a crossing. Each cycle of $\sigma$ corresponds to a self-crossing of the projected curve, each cycle of $\alpha$ corresponds to a part of the curve between two successive crossings, each cycle of $\phi$ corresponds to one of the regions of the plane enclosed by the projected curve. The value $\tau(u)$ is equal to $1$ (resp. $-1$) if the corresponding part of the curve is above (resp. below) the crossing.

In order to transform the projection of the $3$-colour line to this encoding efficiently, one may observe that only the segments of the $3$-colour line on the same column $[k,k+1]\times[l,l+1]\times \setR$ may produce a crossing. If two segments are projected onto intersecting segments, we add a crossing to the knot with the correct values of $\sigma$, $\alpha$, $\phi$ and $\tau$ on the four elements emerging from the crossings. This is a bit tricky algorithmically in order to do it efficiently without too many relabelling of the crossings but this can be done with enough carefulness.

\paragraph{Simplification with Reidemeister moves.}
This paragraph concerns only a numerical optimization so we do not spend lines to describe Reidemeister moves in details and encourage the interested reader to consult for example \cite{lickorish}. 

Reidemeister moves can be easily detected using the previous encoding and we simplify the random knot we produce in order to make the subsequent computations a bit quicker. Only Reidemeister moves of types I and II reduce the number of crossings of a knot; the simplification with Reidemeister moves of type I (resp. II) is done by scanning once the set $H_n$ and simplify the corresponding Reidemeister move before going on with the scan.

Simplifying a Reidemeister move may create new configurations that may be further simplified but we chose not to purge them since they become too rare. We chose instead to \emph{shake} the knot using Reidemeister moves of type III: we scan $H_n$ and, every time a local configuration that can be changed is detected, we apply the Reidemeister move of type III with a probability $p=1/2$ and go on with the scan. Then, we start again with a scan-and-simplify of Reidemeister moves of type I and II. We perform Reidemeister III shakes a number of times chosen by us for each size, in order to obtain a good compromise between the time spent in this simplification and the time gained in the computation of the invariants.

\begin{figure}
\begin{center}
\begin{tikzpicture}
\begin{scope}[xscale =0.75]
\draw[thick] (-1.9,1.85).. controls (-1.5,1.) and (-1,0.5) .. (0,0)
		    .. controls (1,-0.5) and (3.5,-1) .. (4,0)
			.. controls (4.5,1) and (3.5,1.5) .. (2.1,1.95);
			
\draw[thick] (1.9,2.05) .. controls (0.5,2.5) and (-0.5,2.5) .. (-2,2) 
			.. controls (-3.5,1.5) and (-4.5,1.) .. (-4,0) 
			.. controls (-3.5,-1) and (-1,-0.5) .. (-0.1,-0.05);
			 
\draw[thick] (0.1,0.05)	.. controls (1,0.5) and (1.5,1.) .. (2,2) 
			.. controls (2.5,3) and (1,4) .. (0,4) 
			.. controls (-1,4) and (-2.5,3) .. (-2.1,2.15);
			
\draw[ultra thick,->] (0.1,-0.05) -- node [below,red] {$1$} (0.5,-0.25);
\draw[ultra thick,->] (-0.1,0.05) -- node [above,red] {$3$} (-0.5,0.25);
\draw[ultra thick,->] (-0.1,-0.05) -- node [below,red] {$4$} (-0.5,-0.25);
\draw[ultra thick,->] (0.1,0.05) -- node [above,red] {$2$} (0.5,0.25);

\draw[ultra thick,->] (1.9,2.05) -- node [above left,red] {$7$} (1.4,2.2);
\draw[ultra thick,->] (2.1,1.95) -- node [right,red] {$5$} (2.5,1.8);
\draw[ultra thick,->] (2.05,2.10) -- node [above right,red] {$6$} (2.2,2.5);
\draw[ultra thick,->] (1.95,1.9) -- node [below right,red] {$8$} (1.75,1.5); 

\draw[ultra thick,->] (-1.9,2.05) -- node [above right,red] {$9$} (-1.4,2.2);
\draw[ultra thick,->] (-2.1,1.95) -- node [below left,red] {$11$} (-2.5,1.8);
\draw[ultra thick,->] (-2.05,2.10) -- node [above left,red] {$10$} (-2.2,2.5);
\draw[ultra thick,->] (-1.95,1.9) -- node [below right,red] {$12$} (-1.75,1.5); 
\end{scope}
\end{tikzpicture}
\begin{tabular}{|c|c|c|c|c|c|c|c|c|c|c|c|c|}
\hline
$x\in H_n$ & $1$ & $2$ & $3$ & $4$ & $5$ & $6$ & $7$ & $8$ & $9$ & $10$ & $11$ & $12$
\\ \hline 
$\sigma(x)$ &  $2$ & $3$ & $4$ & $1$ & $6$ & $7$ & $8$ & $5$ & $10$ & $11$ & $12$ & $9$
\\ \hline 
$\alpha(x)$ &  $5$ & $8$ & $12$ & $11$ & $1$ & $10$ & $9$ & $2$ & $7$ & $6$ & $4$ & $3$
\\ \hline
$\phi(x)$ &  $8$ & $7$ & $11$ & $10$ & $4$ & $9$ & $12$ & $1$ & $6$ & $5$ & $3$ & $2$
\\ \hline
$\sigma^2\circ\alpha(x)$ &  $7$ & $6$ & $10$ & $9$ & $3$ & $12$ & $11$ & $4$ & $5$ & $8$ & $2$ & $1$
\\ \hline
$\tau(x)$ &  $1$ & $-1$ & $1$ & $-1$ & $-1$ & $1$ & $-1$ & $1$ & $1$ & $-1$ & $1$ & $-1$
\\ \hline
\end{tabular}
\end{center}
\caption{\label{fig:knotencoding}Encoding of a knot as in definition \ref{def:knotencoding} with three permutations $\sigma$, $\alpha$ and $\phi$ and a function $\tau$.}
\end{figure}

\paragraph{Computation of the invariants.} The Alexander polynomial $\Delta_K(t)$ associated to a knot $K$ is a Laurent polynomial with integer coefficients and invariant under symmetry $t\to t^{-1}$. If a knot $K$ can be decomposed as a product of two knots $K_1$ and $K_2$, then $\Delta_K(t)=\Delta_{K_1}(t)\Delta_{K_2}(t)$. It was introduced for the first time in \cite{alexander}.

Many knots have the same Alexander polynomial so the knowledge of this sole invariant is not enough to conclude that two knots are equal. However, this may give some hints, which, together with additional symmetry considerations, helped us to establish some of the conjectures of the last section. 

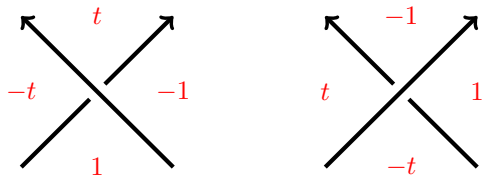
\begin{figure}
\begin{center}
\begin{tikzpicture}
\begin{scope}
\draw[ultra thick,->] (1,-1) -- (-1,1);
\draw[ultra thick,] (-1,-1) -- (-0.1,-0.1); \draw[ultra thick,->] (0.1,0.1) -- (1,1);
\node at (0,1) [red] {$t$};
\node at (-1,0) [red] {$-t$};
\node at (0,-1) [red] {$1$};
\node at (1,0) [red] {$-1$};
\end{scope}
\begin{scope}[xshift=4cm]
\draw[ultra thick,->] (-1,-1) -- (1,1);
\draw[ultra thick] (1,-1) -- (0.1,-0.1); \draw[ultra thick,->] (-0.1,0.1) -- (-1,1);
\node at (0,1) [red] {$-1$};
\node at (-1,0) [red] {$t$};
\node at (0,-1) [red] {$-t$};
\node at (1,0) [red] {$1$};
\end{scope}
\end{tikzpicture}
\end{center}
\caption{\label{fig:Alexandermatrixconvention}Coefficients of the Alexander matrix of a knot depending on the relative positions of a region with respect to a crossing. The arrow corresponds to the direction obtained by following $\sigma^2\circ\alpha$.}
\end{figure}

The advantage of the Alexander polynomial over other invariants is that its evaluations at complex points $t\in\setC$ can be performed efficiently (polynomial time in the number of crossings). We now describe briefly its computation. The projected curve of a knot gives rise to $n$ crossings (cycles of $\sigma$) and $n+2$ regions (cycles of $\phi$). A matrix $M$ with $n$ rows, labelled by crossings, and $n+2$ columns, labelled by regions, is filled in the following way: if the $i$-th crossing is not on the boundary of the $j$-th region, the matrix coefficient $M_{ij}$ is set to $0$; if it is on the boundary, the matrix coefficient $M_{ij}$ is set to $1$, $-1$, $t$ or $-t$ depending on the relative position of the region with respect to the crossing as illustrated in figure~\ref{fig:Alexandermatrixconvention}. If the same region appears several times around a crossing, then the coefficients are added.

Two arbitrary columns corresponding to neighbour regions are then erased from the matrix $M$ in order to produce a square matrix whose determinant is the Alexander polynomial $\Delta_K(t)$ of the knot $K$ up to a multiplicative factor $\pm t^k$, which depends on the choice of the two columns that are erased.

In order to study a large number of knots, we chose to restrict ourselves to the numerical computation of the two integers $|\Delta_K(-1)|$ and $|\Delta_K(i)|$. This avoids the determination of the factor $\pm t^k$ and the divisibility properties of the two integers give strong hints about the prime decomposition of the knot.

\subsection{Numerical programming tools.}

\paragraph{Language and libraries.} All the programs are written in pure \textsf{C++}. Random numbers are produced using the standard \textsf{std::random} library of \textsf{C++}. Compilation is performed by the \textsf{g++} compiler with the standard optimization flags.

The Alexander polynomial computation as a determinant uses the \textsf{Eigen3} library (\url{http://eigen.tuxfamily.org/index.php?title=Main_Page}) for sparse matrices and the absolute value and the logarithm of the determinant of the matrix are computed from the LU decomposition of the sparse matrix where the coefficients are stored as double precision real or complex numbers. 

Parallelization is used only for the Monte-Carlo part: independent copies of the coloured cube are generated by the various processors in order to increase the number of generated knots. The parallelization step is performed by \textsf{OpenMP} directives.

\paragraph{Integer vs. real numbers; determinant vs. logarithm of the determinant.}

The incursion outside the integer numbers for the evaluation of the Alexander polynomial for $t\in\{-1,i\}$ is required by the library \textsf{Eigen3} and \emph{a posteriori} justified by the numerical results, which also justify the computation of $\log| \Delta_K(-1)|$ instead of $|\Delta_K(-1)|$.

For large size cubes, the Alexander matrix is huge (several millions, even after the Reidemeister simplifications) and, as seen below, the knots become highly non-trivial with very large values for the two invariants $|\Delta_K(-1)|$ and $|\Delta_K(i)|$. A major worry is the numerical precision of the values we obtain. In order to check the validity of the real values computed by our program, we performed the following verifications:
\begin{itemize}
\item for small sizes and a large number of knots or for large sizes and a small number of knots, we perform both the calculations in double precision and in multi-precision arithmetic (using the library \textsf{GMP})---~for which the result is exact but the execution very slow~--- and observe that both results agree for $\log |\Delta_K(-1)|$.
\item The computation of $|\Delta_K(-1)|$ (resp. $\log|\Delta_K(-1)|$) is done by multiplication (resp. sum) of the diagonal coefficients of the upper triangular matrix of the LU decomposition of the Alexander matrix. Introducing some additional lines in the open-source \textsf{Eigen3} library, we observe that, even for the size $N=100$ and very large knots, the diagonal coefficients have orders of magnitude between $10^{-2}$ and $10^2$ and thus remain small: the large values of the invariant are produced by a large number of non-large numbers and it justifies the sole use of the logarithm to avoid overflows and gain stability.
\item  for all sizes, we also computed the determinant in double precision but converted the result to the nearest integer when no overflow is reached. The distance between the result and the nearest integer is stored and we verify that it never goes above $10^{-2}$ for sizes up to 40 even for very large values of the invariants.
\end{itemize}

\paragraph{Hardware, memory and computation times of the simulations.}

We ran our simulations on machines with 64 cores Intel Xeon E5-4620 $2.20$~GHz and 256~Go of RAM.

The main limitation is concentrated in the computation of the invariants via LU decomposition. For the size $N=120$, every computation of an invariant requires in average $20$ giga-octets of RAM. For small sizes we ran 40 simulations in parallel but, for the size $100$, we had to reduce to 10 simulations in parallel.

For size $N=100$, the full time to produce one realization of the percolation, the computation of the combinatorial knot, the Reidemeister simplification and the computation of the two invariants $\log |\Delta_K(-1)|$ and $\log |\Delta_K(i)|$ (in double precision) is equal in average to 130 seconds. The Reidemeister simplification and the computations of the two invariants use in average 125 seconds. 

Both the CPU time and the memory usage are mainly used by the evaluation of the invariant: any improvement in \emph{a priori} simplification of the knot or evaluation of the Alexander polynomial may open the door to the study of much larger sizes.

Evaluation in multi-precision using the \textsf{GMP} library for a size $L=50$ slows down the evaluation by a factor $1000$ and, as expected, it gets worse as the size increases. This prevented us from a full arithmetic study of the knot invariant which was our initial goal. The interested reader may however contact the authors to obtain the data file of $10000$ knots in size $N=50$ in multi-precision.

\subsection{Numerical results and conjectures}

The present subsection presents the results of our simulations. In order to simplify the presentation of the results, we write for any real random variable $Z$ its normalized version with a hat:\[
\ha{Z}=\frac{Z}{\Esp{Z}}.
\]

\begin{figure}
\begin{center}
\input{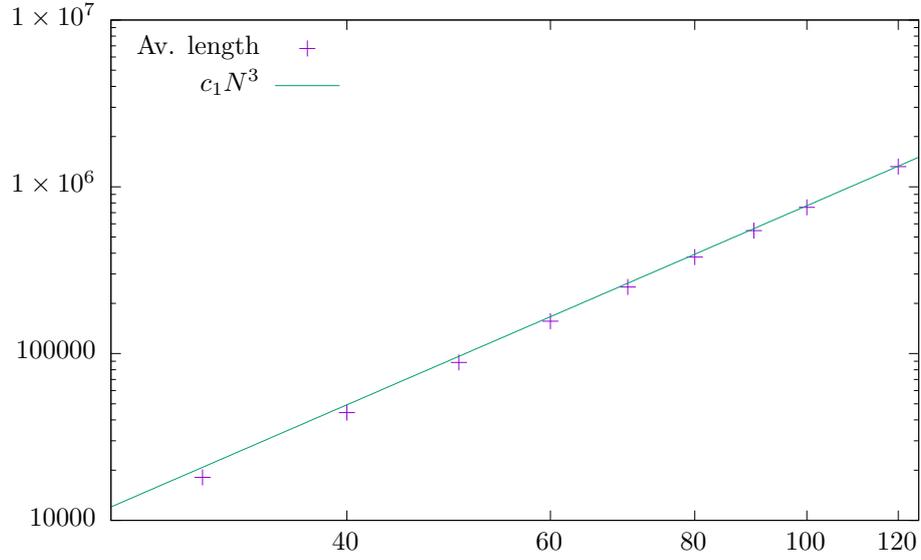}
\end{center}
%\comment{METTRE EN ECHELLE DOUBLEMENT LOGARITHMIQUE AVEC L'ASYMPTOTE $0.76 N^3$.}
\caption{\label{fig:fractaldim} Average length $L_N$ of the interface curve that gives the knot as a function of the size $N$ of the cube. Averages are performed over 100000 independent realizations for each size ranging from $N=30$ to $N=120$. The straight line corresponds to the conjectured $N^3$ scaling.
}
\end{figure}

\begin{figure}
\begin{center}
\input{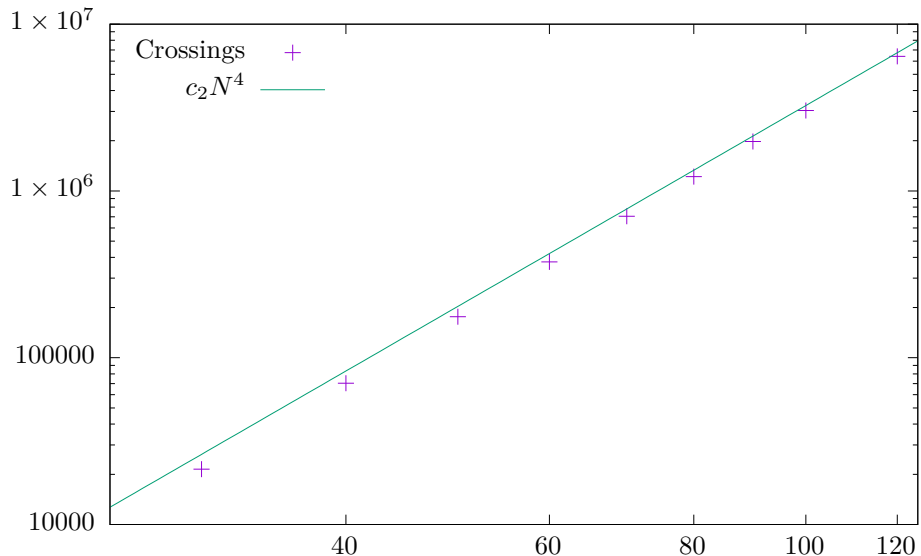}
\end{center}
%\comment{METTRE EN ECHELLE DOUBLEMENT LOGARITHMIQUE AVEC L'ASYMPTOTE $C N^4$.}
\caption{\label{fig:crossings} Average number of crossings of the projection of the interface curve on the xy-plane as a function of the size $N$ of the cube. Averages are performed over 100000 independent realizations for each size ranging from $N=30$ to $N=120$. The straight line corresponds to the conjectured $N^4$ scaling.
}
\end{figure}

\begin{figure}
\begin{center}
\input{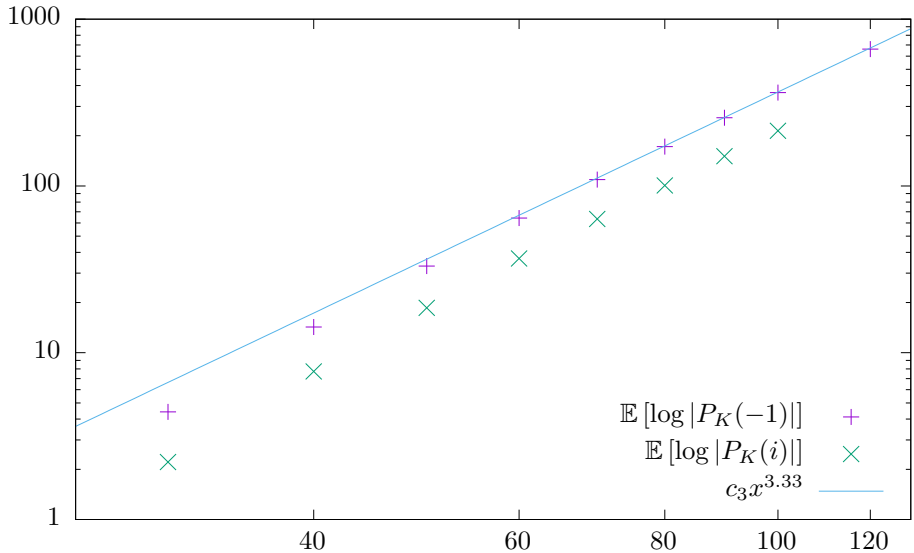}
\end{center}
\caption{\label{fig:logAlex} 
Average values of $\log |\Delta_K(-1)|$ and $\log |\Delta_K(i)|$ as a function of the size of the cube. Averages are performed over 100000 independent knots for each size. The straight line corresponds to a scaling $N^{3.33}$.}
\end{figure}

\paragraph{Error bars.} For every size, we obtained 100000 independant samples for random variables which appear to be square integrable and whose limit law of $\ha{Z}_N$ is also square integrable, so that asymptotic error bars in figures~\ref{fig:fractaldim}, \ref{fig:crossings}, \ref{fig:logAlex} can be estimated using the central limit theorem estimators, and are of order $10^{-2}$ in logarithmic scale even if $\Esp{Z_N}$ increases. They are thus smaller than the symbol sizes in logarithmic scale and are not shown.

\paragraph{Length of the curve.}

The first natural property of the interface curve $\Gamma$ joining the opposite corners of the cube is the length of the curve $L_N$, defined as the number of tetrahedra that are crossed. Figure~\ref{fig:fractaldim} presents the average length of the curve $\Esp{L_N}$ with respect to the size $N$ of the cube. It tends to show that the average length of the curve grows as $N^3$ and thus the curve is space-filling.

\begin{figure}
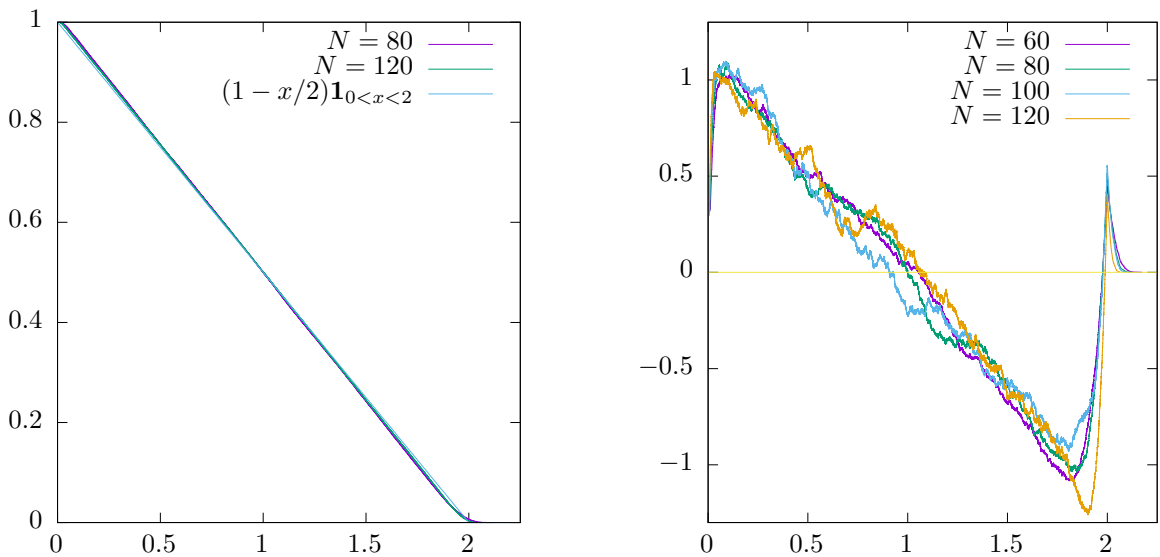

\begin{center}
\input{Lhisto.tex}\hfill \input{Lhisto_finitesize.tex}
\end{center}
\caption{\label{fig:Ldistribution} Empirical repartition functions $G_X(x) = \prob{X> x}$ of the normalized length $\ha{L}_N$ of the interface curve and the conjectured $f(x)=(1-x/2)\indic{x<2}$ limit repartition function are presented on the left graph. Finite size effects corresponding to $N(G_{\ha{L}_N}(x)-f(x))$ are presented on the right graph. The empirical repartition function for each size is calculated using $100000$ independent knots.}
\end{figure}

A more surprising result is the distribution of the length of the interface curve normalized by its empirical mean, as presented in figure~\ref{fig:Ldistribution}. Up to finite size effects around $0$ and $2$, which tend to get resorbed as the size increases, the repartition function tends to converge to the one of a very simple law and it encourages us to formulate the following conjecture.
\begin{conjecture}
The normalized (by its average) length $\ha{L}_N$ of the interface curve $\Gamma$ converges in law to a uniform law on $[0,2]$.
\end{conjecture}
Finite size effects, as presented in figure~\ref{fig:Ldistribution} too, tend to confirm this conjecture since the difference between the empirical distributions and the conjectured one seems to scale as $N^{-1}$.

Both the $N^3$ scaling and the conjectured uniform law already appear in the literature for similar models and non-rigorous field theory derivations \cite{nahum1,nahum2,nahum3} (with a lot of other results on similar models) and with similar conjectures for similar models in \cite{goldschmidtetal} with a much more mathematical flavour. More precisely the $N^3$ scaling is interpreted as a curve with fractal dimension $2$ going through the system $N$ times. Our simulations were made before the knowledge of these results so we did not investigate further in this directions and focused on knots: however, this is a good hint that our simulations (for the known non-knot part at least) are in agreement with the previous results on these models.

\paragraph{Number of crossings.}

The projection of the interface curve on the xy-plane produces crossings. Since, for large sizes, the curves appears to visit a finite fraction of the volume of the cube and the height of a column is $N$, one expects naturally that the number of crossings should grow as $N^4$. This behaviour is indeed observed and is presented in figure~\ref{fig:crossings}. A careful joint observation of the length of the curve and the number of crossings of the projection shows a large correlation between both. 

Moreover, the law of the number of crossings normalized by its average also tend to converge as the size $N$ increases.

\paragraph{Probability of observing a knot with the same invariant as the unknot.}

Identifying the unknot requires much more powerful tools, such as Khovanov homology \cite{KHunknot} than just the Alexander polynomial and, up to now, these tools are not suitable for the numerical treatment of knots with a very large number of crossings. However,  one may measure numerically the fraction of knots such that $|\Delta_K(-1)|=1$ and $|\Delta_K(i)|=1$ and this gives an upper bound to fraction of unknots. As shown in figure~\ref{fig:unknots}, this fraction goes to zero as the size becomes large. However, we have not been able to produce enough knots for sizes $N$ that would be large enough to formulate a precise conjecture about the asymptotic behaviour of the fraction of knots such that $|\Delta_K(-1)|=1$ and $|\Delta_K(i)|=1$. We show the numerical results as a table in figure~\ref{fig:unknots} so that the interested reader may compare it to his guess.

We however remark that the resultats are compatible with an exponential decay as for self-avoiding grid walk (see \cite{modelsofRK} for a nice review) and one may expect that similar arguments such as variants of Kesten pattern theorem may apply in the present case. However, the construction of the curve in our model is different enough so that the proof can not be transposed directly.

\begin{figure}
\begin{center}
\begin{tabular}{|c|c|c|c|}
\hline
Size & Samples & Knots $|\Delta_K(-1)|=|\Delta_K(i)|=1$ & Knots $|\Delta_K(-1)|=1$\\
\hline
30 & 100000 & 30501	& 30576 \\
40 & 100000 & 15650 & 15683\\
50 & 100000 & 8982 & 9000 \\
60 & 100000 & 5634 & 5641\\
70 & 100000 & 3739 & 3743\\
80 & 100000 & 2430 & 2431\\
90 & 100000 & 1763 & 1764\\
100 & 100000 & 1210 & 1211\\
120 & 100000 & not estimated & 616 \\
\hline
\end{tabular}
\end{center}
\caption{\label{fig:unknots} Number of knots such that $|\Delta_K(-1)|=1$ and $|\Delta_K(i)|=1$ for various sizes $N$ of the cube (excepted for the size $120$ for which only $|\Delta_K(-1)|$ is calculated). Only these knots may be the trivial knots.}
\end{figure}

\paragraph{Logarithm of the invariants.}

Plots of the empirical averages of $\log|\Delta_K(-1)|$ and $\log|\Delta_K(i)|$ are represented on figure~\ref{fig:logAlex}. Due to  finite size effects that are still large for sizes around $10^2$, our evaluation of the scaling exponents is far from being precise; however, from the results of figure~\ref{fig:logAlex}, one may produce the estimate $\Esp{\log|\Delta_K(-1)|} \sim c_3 N^\gamma$ with $\gamma \simeq 3.33$.

\begin{figure}
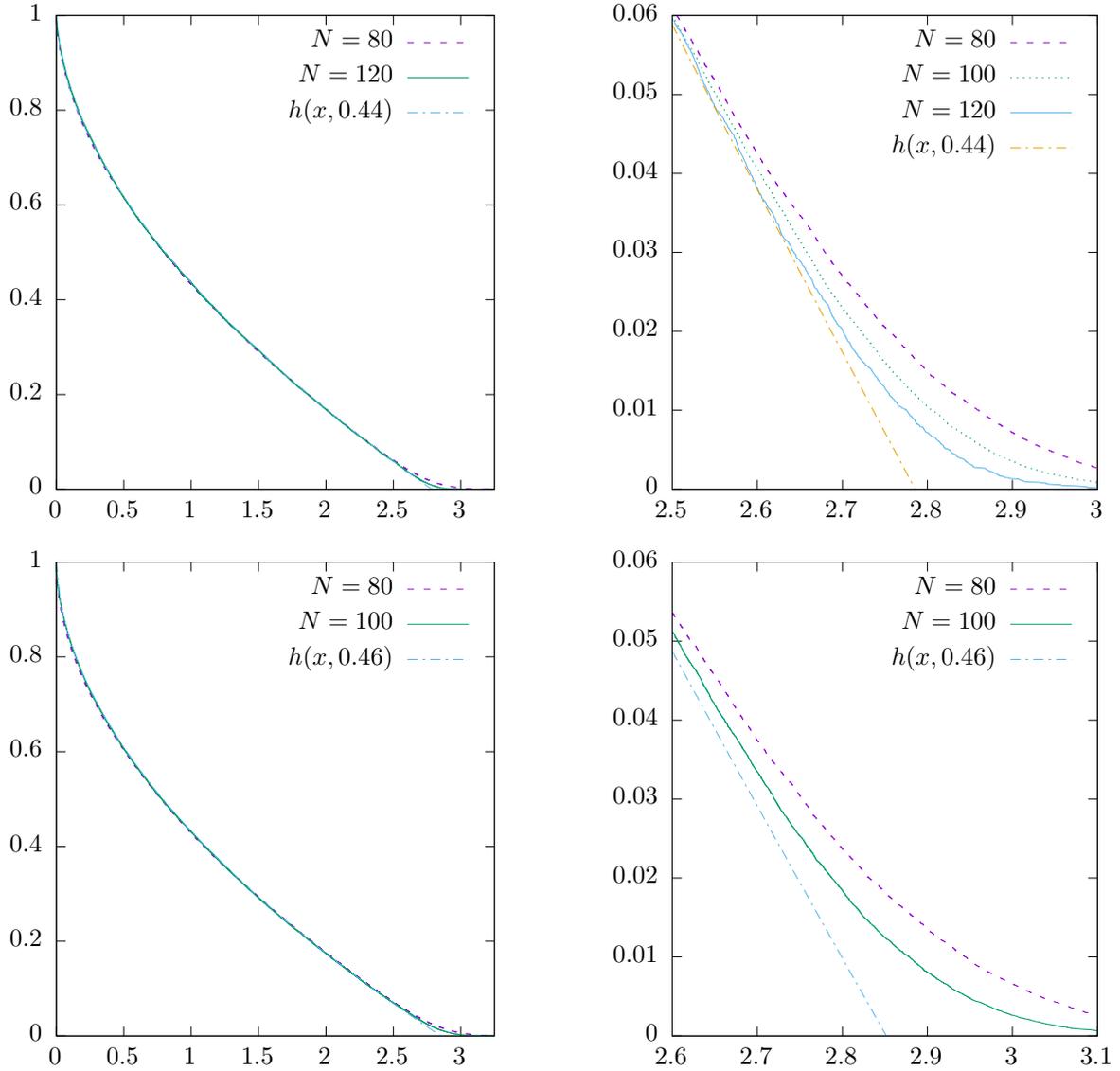

\input{lm2histo.tex}\hfill \input{lm2histozoom.tex}

\input{lm3histo.tex}\hfill \input{lm3histozoom.tex}
\caption{\label{fig:distriinv}Repartition functions $G_X(t)=\Prob(X>t)$ of the normalized random variables $\ha{Y}_N$ (top) and $\ha{Z}_N$ (bottom) corresponding to the logarithm of the Alexander polynomial at $-1$ and $i$ for a cube of size $N=100$. The function $h$ is defined in eq.~\eqref{eq:defreparth}. Plots on the right are zooms of the plots on the left around the cut-off point of $h(x,\alpha)$.}
\end{figure}

Another interesting result is the repartition function $G_X(t)=\prob{X>t}$ of the normalized r.v. $\ha{Y}_N$ and $\ha{Z}_N$ for \begin{align*}
Y_N&=\log(|\Delta_K(-1)|)
\\
Z_N&=\log(|\Delta_K(i)|)
\end{align*}
Results are presented in figure~\ref{fig:distriinv}. One may introduce the repartition function
\begin{equation}
\label{eq:defreparth}
h(x,\alpha) = 1-\min\left( 1, \left(\frac{(1-\alpha)x}{2-\alpha} \right)^{1-\alpha} \right)
\end{equation}
for $x\in\setR_+$ and $\alpha\in (0,1)$. It corresponds to the density $Cx^{-\alpha}$ on an interval $(0,b_\alpha)$ of a r.v. $X$ such that $0\leq X \leq b_\alpha$ a.s. and $\Esp{X}=1$. One observes then very nice fits of the repartition function of $\ha{Y}_N$ by $h(x,\alpha)$ with $\alpha=0.44$ and of $\ha{Z}_N$ by $h(x,\alpha)$ with $\alpha=0.46$. However, finite size effects are too large to decide whether the value $\alpha$ is in fact the same value for both r.v. and lies around $0.45$ or slightly differs. We still dare to formulate the following conjecture.

\begin{conjecture}
For $t=-1$ (resp. $t=i$), there exist numbers $c_t$, $\gamma_t$ and $\alpha_t$ such that the sequence of r.v. $(\log |\Delta_K(t)|)/c_t N^{\gamma_t}$ converge in law to a r.v. with repartition function $h(x,\alpha_t)$.

Approximate values are given by $\gamma_{-1}\simeq 3.33$, $\alpha_{-1}\simeq 0.44$ and $\alpha_{i}\simeq 0.46$.
\end{conjecture}

\paragraph{Divisibility of $|\Delta_K(-1)|$ by invariants of prime knots.}

Due to the computational power of our hardware, we were able to produce only 15000 knots for cube sizes $N=40$ and $N=50$ in multi-precision arithmetic. We have checked that the repartition function of $\log|\Delta_K(-1)|$ obtained this way corresponds (up to finite size effects) to the one obtained in double precision with the same tools as for figures~\ref{fig:logAlex} and \ref{fig:distriinv}. The computation of the invariant $|\Delta_K(-1)|$ as an integer opens the way of its arithmetical study.

Figure~\ref{fig:exactdivisibility} presents our results. Collecting the divisibility of $|\Delta_K(-1)|$ for the first 2977 prime knots contained in \cite{knottable} shows that the divisibility properties of $|\Delta_K(-1)|$ by powers of $3$ and $5$ are close to the ones of random integer numbers. The random knots produced by our simulations exhibit a significant deviation  in their divisibility properties. As the size increases, the divisibility probability of $|\Delta_K(-1)|$ by the numbers $3^k\cdot 5^l$ deviates more and more from uniform random numbers. 

Except if the divisibility properties of prime random knots deviates significantly beyond the first $2977$ prime knots or if our model selects very particular prime knots, our results rather tend to indicate that, as $N$ increases, the random knots are indeed composite and contains more and more prime knots.

\begin{figure}
\begin{center}
\begin{tabular}{|c|c|c|c|c|}
\hline
Divisor $d$ & Freq. $N=40$ & Freq. $N=50$ &
Freq. prime knots & $1/d$ \\
\hline 
$3$ & $0.5418 \mathbf{\;[1.62]}$ & $0.671667 \mathbf{\;[2.02]}$ & $0.364125$&  $0.333333$\\
$5$ & $0.291067 \mathbf{\;[1.46]} $ & $0.370867 \mathbf{\;[1.85]} $ & $0.225059$  &  $0.2$\\
\hline
$9=3^2$ & $0.277933 \mathbf{\;[2.5]} $ & $0.405067 \mathbf{\;[3.65]} $ & $0.152503$ &  $0.111111$\\
$15=3\cdot 5$ & $0.176533 \mathbf{\;[2.6]} $ & $ 0.268067 \mathbf{\;[4.02]}$ & $0.0812899$ &  $0.0666667$\\
$25=5^2$ & $0.0683333 \mathbf{\;[1.71]} $ & $ 0.105867 \mathbf{\;[2.65]}$ & $0.0473631$ &  $0.04$\\
\hline
$27=3^3$ & $0.1224 \mathbf{\;[3.3]} $ & $0.2204  \mathbf{\;[5.95]}$ & $0.0517299$ &  $0.037037$\\
$45=3^2\cdot 5$ & $0.0940667 \mathbf{\;[4.23]} $ & $ 0.165  \mathbf{\;[7.43]}$ & $0.0325831$  &  $0.0222222$\\
$75=3\cdot 5^2$ & $0.044 \mathbf{\;[3.3]} $ & $ 0.079 \mathbf{\;[5.93]} $ & $0.018475(*)$ & $0.0133333$ \\
$125=5^3$ & $0.0156 \mathbf{\;[1.95]} $ & $0.0265333 \mathbf{\;[3.32]} $ & $0.00873362(*)$ & $0.008$ \\
\hline
$81=3^4$ & $0.0505333 \mathbf{\;[4.09]} $ & $0.105267 \mathbf{\;[8.53]} $ & $(*) $ & $0.0123457 $ \\
$135=3^3\cdot 5$    & $0.0422  \mathbf{\;[5.70]}$ & $0.0904667 \mathbf{\;[12.21]} $ & $ (*)$ &$0.00740741 $  \\
$225=3^2\cdot 5^2$  & $0.0246667 \mathbf{\;[5.55]} $ & $0.05 \mathbf{\;[11.25]} $ & $(*) $ &$ 0.00444444$ \\
$375=3\cdot 5^3$  & $0.0107333 \mathbf{\;[4.03]} $ & $0.0198 \mathbf{\;[7.43]} $ & $(*) $ & $0.00266667 $ \\
$625=5^4$ & $0.0028  \mathbf{\;[1.75]}$ & $ 0.0056 \mathbf{\;[3.5]} $ & $(*) $ & $0.0016 $ \\
\hline
$243=3^5$ & $0.0200667 \mathbf{\;[4.87]} $ & $0.0470667  \mathbf{\;[11.44]}$ & $(*) $ & $0.00411523 $ \\
$405=3^4\cdot 5$ & $ 0.0181333\mathbf{\;[7.34]}$& $0.0426667  \mathbf{\;[17.28]}$ & $(*) $ & $0.00246914 $ \\
$675=3^3\cdot 5^2$ & $0.0117333  \mathbf{\;[7.92]}$ & $0.0274  \mathbf{\;[18.50]}$ & $(*) $ & $0.00148148 $ \\
$825=3^2\cdot 5^3$ & $0.0036 \mathbf{\;[2.97]} $ & $ 0.0086 \mathbf{\;[7.10]} $ & $(*) $ & $0.00121212 $ \\
$1875=3\cdot 5^4$ & $0.00193333  \mathbf{\;[3.63]}$ & $0.00413333  \mathbf{\;[7.75]}$ & $(*) $ & $0.000533333 $ \\
$3125=5^5$ & $0.000466667  \mathbf{\;[1.46]}$ & $0.00146667  \mathbf{\;[4.58]}$ & $ (*)$ & $0.00032$\\
\hline
\end{tabular}
\end{center}
\caption{\label{fig:exactdivisibility}Divisibility of $|\Delta_K(-1)|$ by various integers $d$. The first column is the divisor $d$, the second and third column contain the frequency of knots divisible by $d$ for cube sizes $N=40$ and $N=50$. Averages are performed over 15000 independent knots for both sizes. The bold number in parenthesis corresponds to the excess of frequency with respect ot the last column (i.e. is equal to the frequency multiplied by $d$). The fourth column corresponds to the fraction of the first 2977 prime knots (inventoried in the table of prime knots available at \cite{knottable}) with divisibility of $|\Delta_K(-1)|$ by $d$. The last column corresponds to the asymptotic fraction of numbers with $d$ divisibility taken uniformly on $\{1,\ldots,M\}$ for large $M$. Data with $(*)$ are not given or are not reliable since $|\Delta_K(-1)|$ for the first 2977 prime knots have maximal values around $200$.}
\end{figure}

\section{Conclusion}

The present paper generalizes to a three-dimensional geometry the classical construction of interface curves in two-dimensional two-colours percolation, which is known to converge to $SLE_6$ conformally invariant curves. The conformal properties disappear in dimension three for equally probable colours (see below for an additional discussion) but new topological properties, such as knotting, emerge. Our simulations show that the random knots that appear in a large three-dimensional cube with suitable boundary conditions are highly non-trivial:
\begin{itemize}
\item the interface curve becomes space-filling:
\item the frequency of the unknot goes to zero as the size $N$ of the cube increases;
\item $\Esp{\log|\Delta_K(t)|}$, where $\Delta_K$ is the Alexander polynomial and $t\in\{-1,i\}$, diverges with a scaling exponent close to $3.33$ as $N\to\infty$.

\item the divisibility properties of $|\Delta_K(-1)|$, related to the prime knot factorization of the knot, are non-trivial. 
\end{itemize}
It would be very interesting to prove these behaviours rigorously and have exact values for the scaling exponents.

Although we did not manage to reach very large sizes, we have observed also that several laws of r.v., such as the length of the interface curve or the normalized values of $\log|\Delta_K(t)|$, seem to converge to quite simple laws. Up to now, we do not have any hint of explanation for the emergence of these laws. We only hope that our conjectures are true and are not due to artefacts of our simulations: finite size effects on all the figures appear to be significant below the size $N=60$ and much less significant above $N=80$; we have not been able to study carefully finite size effects on the knot invariants so we may only hope that our evaluations of critical exponents $\gamma$ and $\alpha$ are close to the true values.

Besides the mathematical proof (or disproof) of our conjectures, other open questions would be of interest and are of two types: the point of view of knot theory and the point of view of statistical mechanics. 

\paragraph{Knot theory.} The Alexander polynomial of a product knot is the product of the Alexander polynomials of the factor knots. It would be interesting to perform our calculations with multi-precision integers or to compute directly the Alexander polynomial in order to decide if the large knots resemble large prime knots or rather are large product of small prime knots or follow any other intermediate scenario. This would require to use a much larger computation power or to simplify drastically the evaluation of the Alexander polynomial.

There exist invariants that identify the unknot \cite{KHunknot} and some results about the unknot probability for some models of self-avoiding random walks\cite{soterosetal}; their use may help to understand how fast the probability of observing the unknot goes to zero, since our data are not sufficient to formulate a precise conjecture.

If the conjectures are true, it would be interesting to see what they become for other polynomial knot invariants, such as Jones polynomial of HOMFLY polynomial.

\paragraph{Statistical mechanics.} It would be interesting to formulate three-dimensional generalizations of other two-dimensional models from statistical mechanics and to study the knotting properties of level lines. In particular, the three-colour interface curve is a random submanifold of the cube and is the boundary of the bicolour random surface between clusters. Our model adds new topological questions to the study of random submanifolds such as the ones studied in \cite{letendre,gayetwelshinger} or, on the physical side, in \cite{taylordennis}.

Our choice of percolation and crossing curve is very particular: it is the simplest one and it uses as little parameters as possible. However, it may not be the easiest to study and we may imagine very different behaviours of the knot invariants for other models. If one sticks to three-percolation for example, one may use non-equally probable colours and it is known by physicists \cite{bradleyetal,nahum1} that a careful choice of the three colour probability may lead to three-dimensional conformal invariance and thus change drastically the scalings we observed. Further studies in this direction may lead to interessant observations and it will probable require a huge computational power without any optimization on the knot theory size.

\paragraph{Availability of data sets.} Data for all sizes are available on the webpage of the second author or on demand.

%\bibliographystyle{plainnat}
%\bibliography{knots} 

\end{document}